\documentclass[a4paper,11pt]{book}

\usepackage[utf8]{inputenc}
\usepackage{geometry}
\usepackage[hidelinks]{hyperref}
\usepackage{titlesec}
\usepackage{etoolbox}
\usepackage{chngcntr}
\usepackage{graphicx}
\usepackage{lipsum}
\usepackage{xcolor}
\usepackage{amsmath,amssymb,amsfonts}
\usepackage{fancyhdr}
\counterwithout{figure}{chapter}

\patchcmd{\thebibliography}{\chapter*}{\section*}{}{}

\titleformat{\chapter}[frame]{\normalfont}{}{10pt}{\LARGE\bfseries\filcenter}
\begin{document}

\chapter{Patches of developable surfaces bounded by NURBS curves}

\thispagestyle{empty}
\vspace{-0.5cm}
\begin{center}
\begin{large}
		L. 
		Fern\'andez-Jambrina$^\flat$\footnote{leonardo.fernandez@upm.es}
		\\
		\vspace{0.25cm}
		\normalsize ($\flat$)\; ETSI Navales, \\
        \normalsize Universidad Polit\'ecnica de Madrid \\
        \normalsize Avenida de la Memoria 4, 28040-Madrid, Spain.\\ 
\end{large}
\end{center}
\vspace{-0.5cm}

\section{Introduction}


In this talk we review the problem of constructing a developable
surface patch bounded by two rational or NURBS (Non-Uniform Rational
B-spline) curves.  

NURBS curves are curves which are piecewise rational.  That is, they
are a generalisation of spline curves, which are piecewise polynomial
curves.  Similarly we define NURBS surfaces and solids.  NURBS curves
have been the standard in Computer Aided Design \cite{farin} for a 
long time through IGES and STEP specifications.  They are
described by a list of points called control polygon and two list of
numbers: the list of weights and the list of knots.

Developable surfaces are ruled surfaces with null Gaussian curvature
\cite{struik}.  This implies that they can be constructed from planar
surfaces by just cutting, rolling and folding, so that metric
properties such as lengths and angles between curves and areas are
preserved.  These geometric properties are of great interest for steel
and textile industry, since these are pieces designed in the plane and
then combed into space.

For instance, in naval architecture sheets of steel are adapted to fit
into the hull of a ship \cite{kilgore, chalfant, arribas}.  If these sheets are combed just with a
folding machine, the costs are lower than if they require the use of
heat.  In textile industry cloth is planar and is cut and sewn to
produce garments and the quality is improved if it is not stretched 
\cite{rose}. They have also been used for designing facades in 
architecture \cite{architecture} and in automobile industry 
\cite{frey}.

In geometric design the standard relies on the use of rational
B-spline curves and surfaces (NURBS), which are described by control
polygons or nets and lists of weights and knots.  In the case of ruled
surfaces, the control net is formed by just the control polygons of
the bounding curves, since segments can be described by just their
endpoints, which form their control polygon.

This problem has been addressed in several ways 
\cite{computational-line}, but the key 
drawback is that when we require the developable surface to be NURBS 
and bounded by NURBS curves, the possibilities are restricted 
\cite{leonardo-bezier,rational}. 

In some cases, the bounding curves are planar and lie on parallel 
planes \cite{aumann0,maekawa} and this simplifies the problem.

For instance, one can obtain general solutions for developable NURBS
surfaces bounded by NURBS curves
\cite{aumann,aumann1,leonardo-developable,leonardo-elevation, leonardo-triangle}.

Considering the dual space in projective geometry has been also 
profitable, since developable surfaces may be seen as envelopes of 
lines of planes \cite{ravani, pottmann-farin}.  

For this reason our proposal is to consider developable surface
patches which are not NURBS, though bounded by NURBS curves
\cite{develop}.  In fact, we are able to obtain every possible
solution, good or bad, to this problem in our framework.

Our approach is based on performing a reparametrisation on one of the
bounding curves of the surface patch and imposing that the resulting
surface be developable \cite{develop}.  One would expect a
differential equation for the reparametrisation function.  However,
the condition happens to be algebraic.  Another approach
\cite{origami} reduces the null Gaussian curvature condition to
quadratic equations.

Moreover, if the bounding curves are (piecewise)
polynomial or rational of degree $n$, the developability condition is
an algebraic equation of degree $2n-2$.  The degree may be lowered to 
$n-1$ if
the curves are (piecewise) polynomial and lie on parallel planes.

\section{Methods}
We start with a ruled surface parametrised by
$b(t,v)$ and bounded by two parametrised curves, $c(t)$, $d(t)$,
\begin{eqnarray*}b(t,v)=(1-v)c(t)+vd(t), \quad 
t,v\in[0,1].\label{ruled}\end{eqnarray*}

For given $c(t)$ and $d(t)$, this ruled surface will not be
developable in general, but, if it is developable, this feature shall 
not depend on the chosen parametrisation. 

The Gaussian curvature $K(t,v)$ is the quotient of the determinants of
the second, $B(t, v)$, and first, $G( t, v)$, fundamental forms of the
surface patch parametrised by $b(t,v)$, \[K(t,v)=\frac{\det
B(t,v)}{\det
G(t,v)}.\]

The first fundamental form $G( t, v)$ is the usual scalar product
restricted to tangent vectors to the surface at the point $b(t, v)$ 
and is used, for instance, for calculating areas,
\[\mathrm{Area}(S)=\int_{D} \sqrt{\det 
G(t, v)}\,d td v,\] but, since its determinant is positive and appears
at the denominator of $K(u,v)$, we need not pay attention to it, 
though due to Gauss' Theorema Egregium \cite{struik}, the Gaussian 
curvature is an intrinsic property and may be written in terms of 
just the first fundamental form.

On the other hand, the second fundamental form $B(t,v)$ is the extrinsic 
curvature of the surface at the point $b(t, v)$ with $\nu$ as 
unitary normal to the surface and can be constructed with the 
projections of the second derivatives of the parametrisation along the 
unitary normal $\nu(t,v)$ to the surface at $b(t,v)$,
\[   B(t, v)=\left(\begin{array}{cc}
b_{ t t}\cdot   \nu	& b_{ t v}\cdot   \nu	  \\
b_{ v t}\cdot   \nu	& b_{ v v}\cdot   \nu	  \\ 
\end{array}\right)_{( t, v)}.\]

In this sense, the second fundamental form is used to compute the 
normal curvature at points on the surface.

For a ruled surface parametrised as in (\ref{ruled}),
\[\det B( t, v)=\left|\begin{array}{cc}
\left((1-v)c''(t)+vd''(t)\right)\cdot   \nu	& \left(d'(t)-c'(t)\right)\cdot   \nu	  \\
\left(d'(t)-c'(t)\right)\cdot   \nu	& 0	  \\ 
\end{array}\right|_{( t, 
v)}=-\left(\left(d'(t)-c'(t)\right)\cdot\nu\right)^{2},\]
we see that the Gaussian curvature is always non-positive and hence 
there are no elliptic points on a ruled surface, since $\det B(t,v)$ is 
always positive.

Moreover, vanishing Gaussian curvature is equivalent to vanishing 
$\left(d'(t)-c'(t)\right)\cdot\nu$ on the points of the ruled 
surface. 

Since at a point $b(t,v)$ on the surface we have
$b_{t}(u,v)=(1-v)c'(t)+d'(t)$, $b_{v}(t,v)=d(t)-c(t)$ as two tangent
vectors to the surface, we may construct a normal vector to the
surface at $b(t,v)$ as $N=b_{t}\times b_{v}$, and then having a
vanishing Gaussian curvature is equivalent to having a vanishing
triple product, 
\begin{equation}c'(t)\cdot d'(t)\times(d(t)-c(t))=\det\left(c'(t), d'(t),
d(t)-c(t)\right)= 0, \qquad
t\in[0,1]\label{developability}\end{equation} at all points of the
surface patch.
\begin{figure}
\centering
\includegraphics[width=0.3\textwidth]{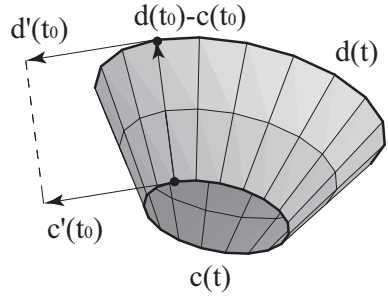}
\caption{Developability is equivalent to having the same tangent 
plane along each ruling of a ruled surface}
\label{tangent}
\end{figure}

As we see in Fig~\ref{tangent}, this is equivalent to requiring that 
the three vectors $c'(t)$, $d'(t)$ and $d(t)-c(t)$ lie on the same 
plane for all values of $t$. This also implies that the normal to the 
surface is the same for all points on the segment (ruling) linking 
$c(t)$ and $d(t)$.


Our contribution to deal with this problem is
based on reparametrisation of one of the bounding curves by a function $T(t)$,
\begin{equation}\label{repar}\tilde 
	b(t,v)=(1-v)c(t)+vd(T(t))\end{equation} and require $\tilde  b(t,v)$ to 
satisfy the null Gaussian curvature condition. 

\section{Results}

The developability condition  (\ref{developability}) applied to 
parametrisations such as (\ref{repar}) can be
seen to be algebraic in $T(t)$, since the dependence on the derivative
$T'(t)$ is factored out by the determinant,
\begin{equation} \det\left(c'(t), \dot d(T), 
d(T)-c(t)\right)= 0,\label{repar1}\end{equation}
where the dot stands for derivation with respect to $T$.

In the case of (piecewise) polynomial or rational curves $c(t)$, 
$d(t)$, further consequences may be derived:

\noindent\textbf{Theorem 1:} Let $c(t)$, $d(T)$, $t,T\in[0,1]$ be rational
curves of degree $n$. 
The parameterized ruled surface,
\[b(t,v)=(1-v)c(t)+vd(T(t)),\qquad t,v\in[0,1],\]
is a developable surface if the reparameterization function $T(t)$ 
satisfies the algebraic equation
\[\det\left(c'(t), \dot d(T), 
d(T)-c(t)\right)_{T=T(t)}= 0,\]
and is a real monotonically increasing function of $t$.

\begin{figure}
\centering
\includegraphics[width=0.25\textwidth]{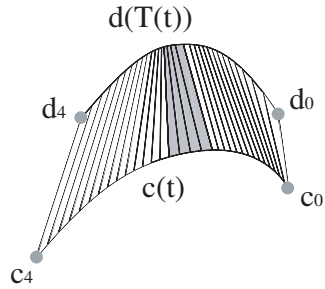}
\caption{Developable patch bounded by two cubic spline curves\label{spline}}	
\end{figure}

This equation is of degree $2n-2$ at most.  If both curves are
(piecewise) polynomial and lie on parallel planes, the equation is of degree $n-1$
at most. We may see an example in Fig~\ref{spline}.	

 
The price to pay is that solutions of this algebraic equation will not
be rational or polynomial in general and $\tilde b(t,v)$ will no
longer be NURBS.

Since the condition on the reparametrisation is algebraic, the number
of possible solutions is finite, but not all of them are geometrically
acceptable. 
\begin{figure}
\centering
\includegraphics[width=0.25\textwidth]{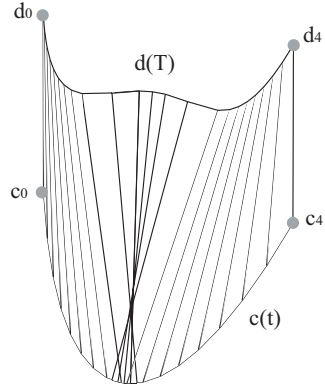}
\caption{Developable surface with a regression area\label{regress}}	
\end{figure}

For being a reparametrisation, $T(t)$ must be a
monotonically increasing function. Otherwise, we would have 
unpleasant regression areas with more than on ruling through some 
points of the curves (See Fig~\ref{regress}). This can be checked with the help
of
\[
T'(t)=\left.\frac{\det\left(c''(t), \dot d(T), 
d(T)-c(t)\right)}{\det\left( \ddot d(T),c'(t), 
d(T)-c(t)\right)}\right|_{T=T(t)},\]
which we derive from the null Gaussian curvature condition.

This implies that monotonicity is granted if
\[\mathrm{sgn}\left(c''(t)\cdot\nu(t)\right)=
\mathrm{sgn}\left.\left(\ddot
d(T)\cdot\nu(t)\right)\right|_{T=T(t)},\] where $\nu(t)$ is the
unitary normal to the ruled surface along the segment at $t$.  This
means that the normal curvatures of both
curves must have the same sign for each value of $t$.  

Hence, acceptable solutions just appear if both curves are
qualitatively similar regarding their curvature.

\noindent\textbf{Theorem 2:} Let $c(t)$, $d(T)$, $t,T\in[0,1]$ be parameterized
curves. Let $T(t)$ be a reparameterization function so that 
\[b(t,v)=(1-v)c(t)+vd(T(t)),\qquad t,v\in[0,1],\]
is a developable surface. 
$T(t)$ is a monotonically increasing function if  and only if for all 
$t$,
\[\mathrm{sgn}\left(c''(t)\cdot\nu(t)\right)=
\mathrm{sgn}\left.\left(\ddot 
d(T)\cdot\nu(t)\right)\right|_{T=T(t)},\] where $\nu(t)$ is the 
unitary normal to the surface along the ruling at $t$.

Or equivalently, for the normal curvatures $k_{n,c}$, $k_{n,d}$ of 
both curves\[\mathrm{sgn}\left(k_{n,c}(t)\right)=
\mathrm{sgn}\left(k_{n,d}(T(t)\right),\] for all values of $t$. 

In the case of parameterizations of class $C^k$ of differentiability, $T(t)$
is of class $C^{k-1}$.

\section{Conclusions}
We have produced a new approach for dealing with the problem of 
constructing a developable surface patch between two parametrised 
curves $c(t)$ and $d(t)$.

This approach is grounded on performing a reparametrisation of one of 
the curves and the developability condition is not a differential 
equation, but an algebraic equation, and provides all possible 
solutions to the problem.

In the case of (piecewise) polynomial or rational curves of degree 
$n$, the developability condition is an algebraic equation of degree 
$2n-2$. Since the most usual degree in Computer Aided Design is 
three, this means we are dealing with a fourth degree equation, 
which can be handled either by numerical or analytical methods.

For (piecewise) polynomial curves of degree $n$, lying on parallel 
planes, the algebraic equation is of degree $n-1$.

Requiring that the reparametrisation function be a monotonically 
increasing function, in order to avoid regression areas on the 
developable surface patch, is achieved if the sign of the normal 
curvatures of both bounding curves is the same at the endpoints of 
each ruling.

With this approach, it is easy to control the final class of 
differentiability of the surface in the piecewise case.

\begin{small}
	
\end{small}

\end{document}